\documentclass[conference]{IEEEtran}\usepackage[]{graphicx}\usepackage[]{color}
\makeatletter
\def\maxwidth{ %
  \ifdim\Gin@nat@width>\linewidth
    \linewidth
  \else
    \Gin@nat@width
  \fi
}
\makeatother

\definecolor{fgcolor}{rgb}{0.345, 0.345, 0.345}

\usepackage{framed}
\makeatletter
 {\par\unskip\endMakeFramed%
 \at@end@of@kframe}
\makeatother

\definecolor{shadecolor}{rgb}{.97, .97, .97}
\definecolor{messagecolor}{rgb}{0, 0, 0}
\definecolor{warningcolor}{rgb}{1, 0, 1}
\definecolor{errorcolor}{rgb}{1, 0, 0}
\newenvironment{knitrout}{}{} %

\usepackage{alltt}
\IEEEoverridecommandlockouts
\usepackage{xspace}
\usepackage{cite}
\usepackage{amsmath,amssymb,amsfonts}
\usepackage{algorithmic}
\usepackage{graphicx}
\usepackage{textcomp}
\usepackage{xcolor}
\def\BibTeX{{\rm B\kern-.05em{\sc i\kern-.025em b}\kern-.08em
    T\kern-.1667em\lower.7ex\hbox{E}\kern-.125emX}}

\usepackage{tikz}
\usetikzlibrary{positioning}

 \usepackage{multirow}

\usepackage[shortcuts]{extdash}
\usepackage[nolist]{acronym}
\usepackage{hyperref} %
\usepackage{units} %
\usepackage{tabularx} %
\usepackage{booktabs}

\newcommand{\ie}{i.\,e.}

\newcommand{\eg}{e.\,g.}

\newcommand{\cf}{cf.}

\newcommand{\afblock}[1]{\noindent{\textbf{#1 }}}
\newcommand{\takeaway}[1]{\noindent{\textit{\textbf{Takeaway:}}} \textit{#1}}

\IfFileExists{upquote.sty}{\usepackage{upquote}}{}
\begin{document}

\begin{acronym}
\acro{AAC}{Advanced Audio Coding}
\acro{ACR}{Absolute Category Rating}

\acro{AoD}{Audio-on-Demand}
\acro{HP}{High Performance}
\acro{LP}{Low Performance}

\acro{NPS}{Net Promoter Score}
\acro{MOS}{Mean Opinion Score}
\acro{QoE}{Quality of Experience}

\acro{SCS}{Short Conversation Scenario}
\acro{WF}{window function}
\acro{LW}{linear window}
\acro{VoD}{Video-on-Demand}
\acro{RMSD}{Root-mean-square deviation}
\end{acronym}

\renewcommand{\sectionautorefname}{Section}

\title{Multi-episodic Perceived Quality\\of an Audio-on-Demand Service
}

\author{
    \IEEEauthorblockN{Dennis Guse\IEEEauthorrefmark{1}, Oliver Hohlfeld\IEEEauthorrefmark{2}, Anna Wunderlich\IEEEauthorrefmark{1}, Benjamin Weiss\IEEEauthorrefmark{1}, Sebastian Möller\IEEEauthorrefmark{1}}
    \IEEEauthorblockA{\IEEEauthorrefmark{1}Technische Universtät Berlin \hspace{2cm} \IEEEauthorrefmark{2}Brandenburg University of Technology}%
}

\IEEEoverridecommandlockouts 

\IEEEpubid{\makebox[\columnwidth]{978-1-7281-5965-2/20/\$31.00 \copyright 2020 IEEE \hfill} \hspace{\columnsep}\makebox[\columnwidth]{ }}
\maketitle

\begin{abstract}
\acl{QoE} is traditionally evaluated by using short stimuli usually representing parts or \emph{single} usage episodes.
This opens the question on how the overall service perception involving \emph{multiple} usage episodes can be evaluated---a question of high practical relevance to service operators.
Despite initial research on this challenging aspect of multi-episodic perceived quality, the question of the underlying quality formation processes and its factors are still to be discovered.

We present a multi-episodic experiment of an \acl{AoD} service over a usage period of 6~days with 93 participants.
Our work directly extends prior work investigating the impact of time between usage episodes.
The results show similar effects --- also the recency effect is not statistically significant.
In addition, we extend prediction of multi-episodic judgments by accounting for the observed saturation.
\end{abstract}

\begin{tikzpicture}[overlay, remember picture]
\path (current page.north) node (anchor) {};
\node [below=of anchor] {%
2020 Twelfth International Conference on Quality of Multimedia Experience (QoMEX)};
\end{tikzpicture}

\section{Introduction}

Traditional research on perceptual \acf{QoE} investigate short time-scales spanning from several seconds up to several minutes and involving only judgments of single interaction.
For these, it has been shown that later parts of the stimuli as well as the worst performance have a higher impact on post-experience (\ie{}, retrospective) judgments~\cite{weiss_modeling_2009}.
These two effects are denoted as recency effect and peak effect, which are well-known from research on recall (\eg{},~\cite{kahnemann_pain_1993}).
For predicting retrospective \ac{QoE} judgments, it has been shown that a weighted average with a higher weighting on more recent momentary judgments or performance performs sufficiently.
In this regard, \emph{multi-episodic perceived quality} investigates the formation process of a subjective quality impression for a service or system that is used repeatedly.

However, the underlying formation process of perceived quality over several usage episodes (\ie{}, multi-episodic perceived quality) is not well understood---especially considering usage periods of days, months, or even years.
Following~\cite{guse_multiepisodic_2017}, a \emph{usage episode} is defined as \emph{a distinct and self-contained interaction by a user with a service or system to achieve his or her goal(s)}.
Investigating and understanding the formation process of multi-episodic perceived quality is of high practical relevance to service operators as telecommunication services are prone to performance fluctuations (\eg{}, varying network conditions).
These fluctuations may be perceived by a user and therefore affect his/her instantaneous quality as well as episodic and multi-episodic quality.

This especially includes cloud-based multi-media services (\eg{}, \acf{AoD} and \acf{VoD} streaming services that have become popular Internet applications).
Research on multi-episodic perceived quality could show that a recency effect occurs (\eg{}, \cite{guse_multiepisodic_2017, guse_modelling_2014}) as well as a duration neglect \cite{guse_duration_2016}.
Despite these first initial findings, the formation process remains far from being understood---rooted also in the high complexity to perform multi-episodic experiments given their duration and the between-subject design.

In this paper, we aim to better understand this formation process using an \ac{AoD} service.
That is, if effects observed on multi-episodic perceived quality in one session can also be observed when the usage is extended to multiple usage episodes.
Our experiment involves 93~participants using the \ac{AoD} service twice per-day over 6~days.
We complement our study by also by applying the \ac{NPS} to investigate the impact on customer loyalty \cite{reichheld_one_2003}.
While it is of questionable reliability~\cite{npscritism} and not well-established the \ac{QoE} domain, it is popular in marketing and user retention analyses, e.g., as an \ac{AoD} service might apply in practice.

\afblock{Structure.}
We first review related work in~\autoref{sec:rw} on which we base our study.
We describe our hypotheses and research design in~\autoref{sec:methodology} and then discuss the study results in \autoref{sec:results}.
Finally, we conclude the paper and give an outlook on future work towards understanding multi-episodic QoE.

\section{Related Work}\label{sec:rw}
\afblock{Multi-day experiments.}
Research on multi-episodic perceived quality emerged in 2011 with multi-day experiments.
A first experiment evaluated Skype calls performed on a daily basis over 12~days~\cite{moller_single-call_2011}.
Each pair of subjects performed two video telephony calls per day while solving one task per call.
This task-driven approach was selected to create a realistic usage situation as well as a comparable usage behavior (\ie{}, one \ac{SCS}~\cite{itu-t_recommendation_p.805_subjective_2007}).
\emph{Episodic judgments} (\ie{}, perceived quality of one usage episode) were directly collected after finishing each call.
\emph{Multi-episodic judgments} (\ie{}, all so far experienced usage episodes) were collected after the 2nd, 7th, and 12th~day.
Within each multi-episodic condition two performance~levels limiting the overall transmission bandwidth were applied: \ac{HP} and \ac{LP}.
Although the results were rather limited this experiment showed that multi-episodic perceived quality can be assessed successfully in a field experiment by applying a between-subject design---which we therefore adopt in this experiment.
The results show that episodic judgments are reduced for \ac{LP} usage episodes.
Moreover, the results indicate that also subsequent episodic judgments are negatively affected even if these were presented in \ac{HP}.
Interestingly, a slight increase for episodic judgments was observed over the usage period.
With regard to the formation process of multi-episodic judgments the results are rather limited.
This is most likely due to the limited impact of \ac{LP} usage episodes.
Similar results were found for a \ac{VoD} service~\cite{guse_macro-temporal_2013}.
They further observed discrepancies between episodic judgments of \ac{LP} usage episodes and the final multi-episodic judgment.
Precisely, they observed that for a service providing mainly severely \ac{LP} usage episodes, episodic judgments are more positive than multi-episodic judgments.
However, the results are limited by the number of participants as well as that the defined performance levels could not be achieved.

Subsequently, an experiment with a service bundle consisting of an \ac{AoD} service and a \ac{VoD} service over a usage period of 15~days~\cite{guse_macro-temporal_2013} was conducted.
Combining their results with \cite{moller_single-call_2011} and \cite{guse_macro-temporal_2013}, they presented initial models for predicting multi-episodic judgments based upon episodic judgments.
They could show that a \emph{linear moving average} outperforms a \emph{windowed average}.
Accounting for a peak effect resulted in decreased prediction performance.

Overall, the results of multi-episodic perceived quality over usage periods spanning several days is rather limited.
One reason might be that such experiments require to be conducted outside of the laboratory.
Thus, the usage environment is uncontrolled and often an elaborate technical setup necessary.
Also, the required between-subject design increases the effort.

\afblock{Session quality.}
Multi-episodic perceived quality was further investigated in \emph{individual sessions} (\ie{}, continuous use of the same service with multiple usage episodes).
This complements multi-day experiments as it is not yet known if and how the time between usage episodes affects multi-episodic judgments.
In \cite{guse_duration_2016}, an \ac{AoD} service was used to determine if the duration of a \ac{LP} usage~episode affects a subsequent multi-episodic~judgments.
It was observed that the duration of one \ac{LP} usage episode does not affect the episodic and multi-episodic~judgment.
Subsequently, \cite{guse_multiepisodic_2017} conducted two experiments with overall 205~participants to investigate the impact of the usage situation in case of speech telephony.
Both experiments consisted of six usage~episodes that needed to be solved subsequently.
In the first experiment, a pair of participants needed to solve one \ac{SCS} per usage~episode together.
The second experiment, was conducted by each participant alone simulating a 3rd-party listening situation.
Here, recordings of the first experiment were used and participants needed to transcribe all information necessary to solve the \ac{SCS}.
Here, the effects of presenting more \ac{LP} usage episodes as well as presenting more \ac{HP} usage episodes subsequently were investigated.
Most notably, the results indicate that the usage situation has a very limited impact on episodic and multi-episodic judgments.
Increasing the number of \ac{LP} usage episodes resulted in a reduction of the subsequent multi-episodic judgment while remaining well above the episodic judgments of \ac{LP} usage~episodes.
This indicates that previously experienced usage episodes still affect this judgment and that the formation process is not a pure average.
Also a positional impact could be observed in both experiments: increasing the number of \ac{HP} usage~episodes following \ac{LP} usage~episode(s) before a multi-episodic judgment limits the observed reduction.

\takeaway{
Despite first findings, the formation process of multi-episodic perceived quality remains far from being understood.
While initial work in one session found some interesting insights, multi-day experiments so far remained mainly inconclusive.
In this paper, we address this issue by extending \cite{guse_multiepisodic_2017} from one session to multiple days while using a similar experimental design.
}

\section{experiment}\label{sec:methodology}
The goal of this experiment is to investigate if the effects observed on multi-episodic perceived quality in one session can also be observed if the usage period is extended to multiple days.
This complements prior work and enables to improve prediction models for multi-episodic perceived quality.

\subsection{Design}
Our multi-episodic perceived quality experiment also follows a between-subject design (\ie{}, only one multi-episodic condition was presented to each participant) in which participants use an \ac{AoD} service twice per day.
We chose a usage period of 6~days to be able to investigate a higher number of multi-episodic conditions compared to prior work.
To directly embed our experiment into related work, we follow the experimental design of~\cite{guse_multiepisodic_2017}.

We choose an \ac{AoD} service for two reasons.
First, \ac{AoD} is a popular Internet service (\eg{}, offered by popular apps such as Spotify or Audible).
Second, an \ac{AoD} service enables a simple technical setup in which the experiment can be conducted by each participant alone.
That is, no interaction (and thus no coupling) with other participants---as in typical interaction experiments---is needed.
This reduces the experimental complexity, avoids social effects, and reduces the effort to conduct the experiment for the participants.
We used an \emph{audio book} as content to \emph{i)} keep the experiment interesting for participants and \emph{ii)} enable us to verify that the content was consumed.
We chose the audio book \emph{City of the Beasts} from Isabel Allende as it was used in \cite{guse_duration_2016}.
While participants could not chose the content, it limits the effort to prepare the experiment and omits differences in content as source for noise.
This audio book was cut into individual, self-contained parts of~\unit[6..8]{min} length.
One part was presented in each usage episode.
This should enable participants to focus on the content while limiting their effort.
The content was presented in it's chronological order.

In line with prior work, two performance levels \ac{HP} and \ac{LP} were applied.
\ac{HP} denotes the highest performance, yielding only very limited to no perceptible degradations.
\ac{LP} denotes the worst performance and is expected to provide a severely lower perceived quality.
For \ac{HP}, the source material (\textsc{\lowercase{CD}}, \unit[44.1]{kHz}, stereo) was encoded with MP3~(\unit[192]{kbit/s}).
The bitrate was selected to produce no audible impairments.
For \ac{LP}, the content was encoded with \textsc{LPC\=/10}\footnote{The LPC\=/10 encoded content was re-encoded with MP3~(\unit[192]{kbit/s}) for the actual transmission and reproduction.)}.
This codec was also used in prior work (\eg{}, \cite{guse_multiepisodic_2017, guse_duration_2016}) as it provides a severe degradations while providing speech intelligibility.
We remark that the LP encoding is \emph{unrealistic} for any multi-media streaming service.
However, other quality degradation types beyond our scope (e.g., stalling) are expected in practice, yielding also quality fluctuations (LP/HP).
We added the LP encoding as reference to prior work that used the same encoding to study the quality formation process of multi-episodic judgements.
Participants used their own computer and pair of headphones to access the \ac{AoD} service via the Internet using a \textsc{\lowercase{HTML5}}-capable web browser.
The system was implemented using \cite{guse_thefragebogen_2019}.
To exclude Internet-induced artifacts, the audio content was \emph{preloaded} prior to starting each usage episode.

We apply the following hypotheses to evaluate multi-episodic judgments in one session~\cite{guse_multiepisodic_2017}.

\newcommand{\hyponumber}{\textbf{H1}\xspace}
\newcommand{\hypoposition}{\textbf{H2}\xspace}
\newcommand{\hypoconsecutive}{\textbf{H3}\xspace}
\begin{itemize}
  \item (\emph{H1}) increasing the number of \ac{LP} usage episodes leads to a higher reduction in multi-episodic judgments.\label{hypo:number}
  \item (\emph{H2}) presenting \ac{HP} usage episodes after \ac{LP} usage episodes limits the reduction in multi-episodic judgments.\label{hypo:position}
  \item (\emph{H3}) presenting \ac{HP} usage episodes between \ac{LP} usage episodes leads to a higher reduction than presenting the \ac{LP} usage episodes consecutively.\label{hypo:consecutive}
\end{itemize}

Based upon the three hypotheses, seven multi\-/episodic conditions were created.
Performance was only varied between days.
Following~\cite{guse_multiepisodic_2017}, the first three days were presented in \ac{HP} to provide a common baseline for the between-subject design.
\ac{LP} episodes were only presented from the 4...6th~day.

\begin{table}\label{tab:lab:hypothesesComparison}
 \centering
 \caption{Overview on multi-episodic conditions.}
 \label{tab:lab:hypothesesComparison}
 \begin{tabularx}{0.6\columnwidth}{c|c|c|c|c|c}
 \multirow{2}{*}{Condition}& \multicolumn{4}{c}{Episodic performance}        \\
           	& 1-3	& 4			& 5           & 6 \\
 \midrule
 C0         & \ac{HP} 	& \ac{HP}           & \ac{HP}          & \ac{HP} \\
 \hline
 C1         & \ac{HP} 	& \textbf{\ac{LP}}& \ac{HP}          & \ac{HP} \\
 \hline
 C3         & \ac{HP} 	& \ac{HP}          & \ac{HP}          & \textbf{\ac{LP}} \\
 \hline
 C4         & \ac{HP} 	& \textbf{\ac{LP}} & \textbf{\ac{LP}} & \ac{HP} \\
 \hline
 C5         & \ac{HP} 	& \ac{HP}          & \textbf{\ac{LP}} & \textbf{\ac{LP}} \\
 \hline
 C6         & \ac{HP} 	& \textbf{\ac{LP}} & \textbf{\ac{LP}} & \textbf{\ac{LP}} \\
 \hline
 C8         & \ac{HP} 	& \textbf{\ac{LP}} & \ac{HP}          & \textbf{\ac{LP}} \\
 \end{tabularx}
\end{table}

The multi-episodic conditions are shown in \autoref{tab:lab:hypothesesComparison}.
C1 and C3 present either the 4th or the 6th~day in \ac{LP}.
C4, C5, and C8 present two days between the 4th and the 6th~day in \ac{LP}.
C6 presents all usage episodes on these three days in \ac{LP}.
C0~(\ac{HP} only) was explicitly skipped in this experiment as \cite{guse_multiepisodic_2017} did not find evidence that the multi-episodic judgments would be affected.
Also, the effect of a slight increase over the usage period reported by \cite{moller_single-call_2011} was rather small and was observed over a usage period of 14 days.
Therefore, we use the multi-episodic judgment after the 3rd~day as C0.

For the investigation of \hyponumber, the results of C0, C3, C5, and C6 can be compared.
\hypoposition can be evaluated by comparing the results of C1 and C3 as well as C4 and C5.
Finally, \hypoconsecutive can be evaluated by comparing the results of C8 with C3 and C5.

\subsection{Procedure}
\afblock{Introductory session.}
The experiment started with a introductory session.
The goal was to explain the experimental procedure and to collect demographic data.
Subsequently, a short training presenting short stimuli of typical audio degradations was conducted.
Finally, two usage episodes with the \ac{AoD} service needed to be conducted to show participants how to use the service.

\begin{figure}
	\centering
	\includegraphics[width=0.99\columnwidth]{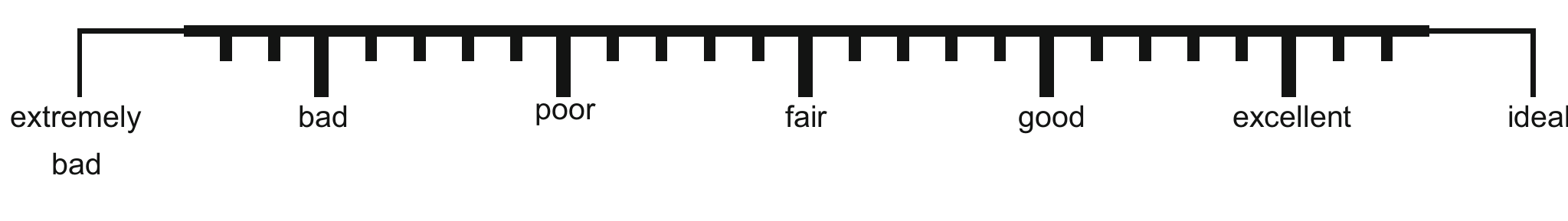}
	\vspace{-1cm}
	\caption{Continuous 7-point scale defined in \cite{itu-t_recommendation_p.851_subjective_2003}.}
	\label{fig:quality_scale}
\end{figure}

\afblock{Experiment.}
The multi-episodic part of this experiment began the day after the introductory session.
Here, the usage episodes needed to be conducted daily between \unit[7]{am} and \unit[1]{pm} as well as \unit[3]{pm} and \unit[10]{pm}, respectively.
In line with prior work, judgments were taken on the 7\=/point continuous category rating scale (see~\autoref{fig:quality_scale}).
Episodic judgments were taken after every usage episode and multi-episodic judgments after the second usage episode of the 3rd~day and the 6th~day.

\afblock{Control questions.}
As participants could not be supervised during the experiment, we presented two content-related questions after every usage episode.
For every questions the correct answer out of three options needed to be selected.
This allows to evaluate if a participant had experienced the content.
This was inspired by cheating prevention approaches for crowdsourcing~\cite{hossfeld2014qoe}.

\afblock{Final assessment.}
On the day after the 6th~day usage period, a debriefing was conducted with every participants.
Here, also the \ac{NPS} was assessed asking how likely it would be that the provided service would be recommended to friends or colleagues (0 \emph{not likely at all} until 10 \emph{extremely likely}).

\subsection{Participants}
The experiment was conducted in Berlin from September until November~2015 with 57~female and 38~male participants aging from of 18~to~33~years ($\mu=25.8$, $\sigma=4.0$). %
Participants were required to have normal hearing capabilities.
Also they needed to comply with the defined schedule and complete all questionnaires.
As a reminder, participants were informed by email, when a usage episode should be conducted.
Successful participation was compensated with \unit[20]{\textsc{\lowercase{EUR}}}.

\section{Results}\label{sec:results}
We next present the results of our experiment.
First, the participants are screened for inconsistent judgments and error rate of the content-related questions is evaluated.
Second, the potential impact of the between-subject design is evaluated.
Then, the multi\-/episodic judgments are evaluated with regard to the three investigated hypotheses.

\subsection{Plausibility Checks}
For the evaluation of consistent judgments, we use the episodic judgments and evaluate these individually.
We consider a participant to be inconsistent if more than two episodic judgments exceed the 1.5~$\times$ \emph{interquartile range} of the performance levels.
None of the participants fulfilled this criteria.
With regard to the content-related questions, we assume that participants should at least answer 50\% correctly.
Otherwise, a participant did not seem to follow the experimental instructions and thus would be excluded from further data evaluation.
Out of the 24~questions, participants answered on average 20.8 questions ($\sigma = 2.7$) correctly.
One participant was excluded. %

\subsection{Between-subject Design}
Given that a between-subject design is applied, we next investigate if this affects the episodic judgments between multi-episodic conditions.
We show the \ac{MOS} per multi-episodic condition in~\autoref{tab:field:e6episodic}).
For episodic judgments of \ac{HP}, a significant difference is found ($H(5)=33.4978$, $p<0.001$).
The post-hoc test shows that C5 is different to all other conditions ($p<0.05$) and that C3 is different to C4 and C8 ($p<0.02$).
For \ac{LP}, differences between conditions are also found ($H(5)=18.2748$, $p=0.0026$).
The post-hoc test shows that C5 is different than C1, C4, and C8 ($p<0.05$).
It must be noted that episodic judgments of C5 resulted in the highest \ac{MOS} for \ac{HP} and in the lowest \ac{MOS} for \ac{LP}.
This is unexpected but not unlikely due to the between-subject design considering the number of multi-episodic conditions.
A detailed analysis did not reveal reason(s) for this and we therefore presume that this does not prevent comparision between multi-episodic conditions.

For the multi\-/episodic judgment after the 3rd~day, no significant differences between conditions are observed ($H(5)=5.2111$, $p=0.3907$).
This indicates that as long as only \ac{HP} episodes were presented, the between-subject design did not affect multi\-/episodic judgments.
\takeaway{The between-subject design did not appear to affect the multi-episodic evaluation.}

\begin{table}
	\centering
	\caption{Number of participants and episodic judgments per multi-episodic conditions. Reported as \ac{MOS} with standard deviation in brackets.}
	\label{tab:field:e6episodic}
    \begin{tabularx}{0.85\columnwidth}{c|c|c|c|c}
	Condition & \ac{LP} days & Participants &  \ac{HP} & \ac{LP} \\
	\midrule
    C0 & - 		& - 	& - \\
	\hline
	C1 & 4 		& 16 & 4.5 (0.9) 	& 1.3 (0.8) \\
	\hline
	C3 & 6 		& 14 & 4.7 (0.7) 	& 1.3 (0.8) \\
	\hline
	C4 & 4..5	& 13 & 4.5 (0.7)		& 1.3 (0.5) \\
	\hline
	C5 & 5..6	& 18& 5.0 (0.8)	& 0.9 (0.5) \\
	\hline
	C6 & 4..6	& 14	& 4.6 (0.6)		& 1.2 (0.7) \\
	\hline
	C8	& 4 and 6	& 15 & 4.5 (0.6)		& 1.2 (0.7) \\
	\end{tabularx}
\end{table}

\subsection{Multi-episodic Judgments}
We next evaluate the three hypotheses stated in Section~\ref{hypo:number} using the final multi-episodic judgments.

\afblock{\hyponumber (increasing number of \ac{LP} usage episodes).}
We first investigate if increasing the number of \ac{LP} episodes \emph{before} a multi\-/episodic judgment results in a decrease of this judgment (\ie{}, a reduction in perceived quality is reported).
This hypothesis is evaluated by comparing C0, C3, C5, and C6 (\ie, 0-3 \ac{LP} usage episodes).
\autoref{tab:field:hyponumber} shows both multi-episodic judgments.
C0, C3, C5, and C6, are significantly different ($H(3)=68.3657$, $p<0.001$).
A post-hoc test finds that C0 is significantly different to all other conditions ($p<0.001$).
Also, C3 and C5 ($p=0.002$) as well as C3 and C6 are significantly different ($p<0.001$).
For C5 and C6, no significant difference is found ($p=0.366$).

\begin{table}
	\centering
	\caption{\hyponumber: multi\-/episodic judgments after the 6th~day. Reported as \ac{MOS} with standard deviation in brackets.}
	\label{tab:field:hyponumber}
	\begin{tabularx}{0.75\columnwidth}{c|c|c}
	Condition	& \ac{LP} episode(s) 	& Multi-episodic judgment\\
	\midrule
	C0          & (\ac{HP} only) & 4.7 (0.6) \\
	\hline
	C3			& 6				& 3.6 (0.6)\\
	\hline
	C5			& 5..6			& 2.5 (1.0)\\
	\hline
	C6			& 4..6			& 2.4 (0.7)\\
	\end{tabularx}
\end{table}

As a result, \hyponumber can only be partly accepted as the multi-episodic judgment decreased, but only for up to two \ac{LP} usage episodes.
The underlying reason for the observed saturation could not be derived from this experiment.
\takeaway{We find that increasing the number of \ac{LP} days directly before the multi\-/episodic judgment negatively affects this judgment.
Here, we observe a reduction of approximately \unit[1]{pt} per \ac{LP} usage episode for up to two episodes.
No further decrease can be observed in case of three \ac{LP} days.
It must be noted that the multi\-/episodic judgment remains \unit[1]{pt} higher than the episodic judgments of \ac{LP} usage episodes.}

\afblock{\hypoposition (increasing number of \ac{HP} usage episodes after \ac{LP}).}
We next evaluate the presence of a recency effect (\ie{}, if presenting \ac{HP} usage episodes after \ac{LP} usage episodes limits the reduction in multi-episodic judgments).
This hypothesis is investigated by comparing C1 vs. C3 (\ie{}, one day \ac{LP}) and C4 vs. C5 (\ie{}, two days \ac{LP}).
\autoref{tab:field:hypoposition} shows the final multi\-/episodic judgment.
With regard to the final multi\-/episodic judgment neither C1 and C3 ($W=138.50$, $p=0.136$, one-sided) nor C4 and C5 ($W=151.00$, $p=0.087$, one-sided) are significantly different.
\takeaway{Unlike prior work which observed a recency effect on shorter usage periods, we did not find clear indications.
As a service provider, \ac{LP} usage episodes should thus be avoided as the subsequent \ac{HP} usage episodes do not make up for prior \ac{LP} experiences.}

\begin{table}[b]
	\centering
	\caption{\hypoposition: multi\-/episodic judgment after the 6th~day for. Reported as \ac{MOS} with standard deviation in brackets.}
	\label{tab:field:hypoposition}
	\begin{tabularx}{0.75\columnwidth}{c|c|c}
	Condition   & \ac{LP} episode(s) 	& Multi-episodic judgment \\
	\midrule
	C1			& 4				& 4.1 (0.7)\\
	\hline
	C3			& 6				& 3.6 (0.6)\\
	\hline
	\hline
	C4			& 4..5			& 3.0 (1.1)\\
	\hline
	C5 			& 5..6			& 2.5 (1.0)\\
	\end{tabularx}
\end{table}

\afblock{\hypoconsecutive (consecutive vs.~non-consecutive \ac{LP} usage episodes).}
Finally, we evaluate if \ac{HP} usage episodes between \ac{LP} usage episodes lead to a higher reduction than presenting the \ac{LP} usage episodes consecutively.
That is, do users prefer performance switches between usage episodes or rather continuous presentation of similar performing usage episodes.
This hypothesis can be evaluated by comparing C4 and C5 with C8.
C4~and C5 present each two days \ac{LP} consecutively, whereas C8 presents the 4th and the 6th~day in~\ac{LP}.
\autoref{tab:field:hypoconsecutive} shows the final multi\-/episodic judgment for these conditions.
These conditions are not significantly different ($H(2)=2.3809$, $p=0.3041$).
As a result, \hypoconsecutive must be rejected.
In fact, the slight improvement in the multi\-/episodic judgment of C8 compared to C5 might be explained by a recency effect.
\takeaway{Performance switches between usage episodes do not seem affect the formation process.}

\begin{table}
	\centering
	\caption{\hypoconsecutive: multi\-/episodic judgment after the 6th~day. Reported as \ac{MOS} with standard deviation in brackets.}
	\label{tab:field:hypoconsecutive}
	\begin{tabularx}{0.75\columnwidth}{c|c|c}
	Condition   & \ac{LP} episode(s) 	& Multi-episodic judgment \\
	\midrule
	C4			& 4..5			& 3.0 (1.1)\\
	\hline
	C8			& 4 and 6	& 2.7 (0.4)\\
	\hline
	C5			& 5..6			& 2.5 (1.0)\\
 \end{tabularx}
\end{table}

\afblock{State of the art.}
Our results are in line with single-session results in prior work that investigated multi-episodic use in one sessions of one hour~\cite{guse_multiepisodic_2017}.
This indicates that the time between usage episodes in the studied experiments might only have a limited impact on multi-episodic judgments.
We infer that it is very beneficial to investigate first multi-episodic perceived quality in one or more sessions of multiple usage episodes each.
Then these findings can be verified and extended in multi-day experiments.
This will also enable to create prediction models for multi-episodic perceived quality.

\subsection{Net Promoter Score}
\begin{figure}
	\centering
\begin{knitrout}
\definecolor{shadecolor}{rgb}{0.969, 0.969, 0.969}\color{fgcolor}
\includegraphics[width=\maxwidth]{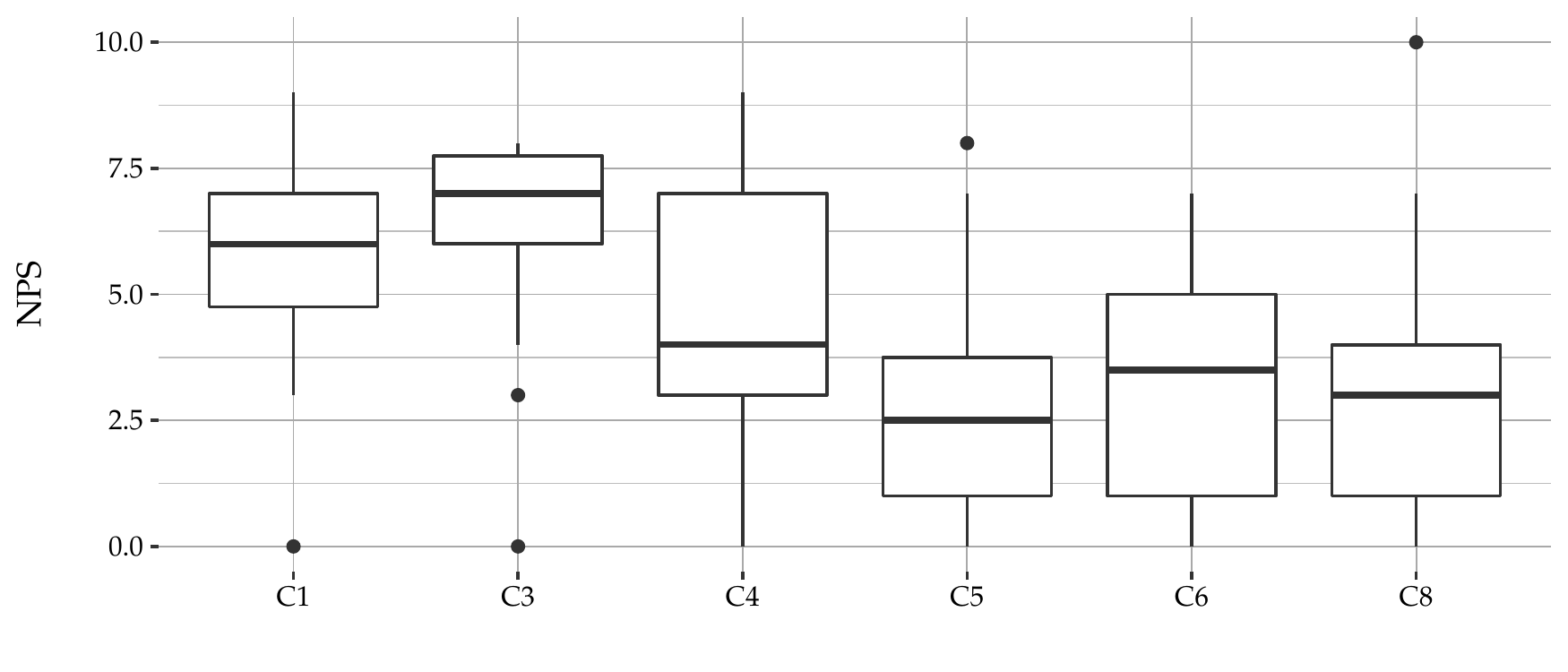} 

\end{knitrout}
	\caption{Boxplot of the \ac{NPS} per multi-episodic condition.}
	\label{fig:pred:E6nps}
\end{figure}

The \ac{NPS} assesses how likely it would be that the provided service would be recommended to friends or colleagues (0~not likely at all to 10~extremely likely).
It thereby divides participants into promoters (\unit[9-10]{pt}), passives (\unit[7-8]{pt}), and detractors (\unit[0-6]{pt}) to determine the growth and churn of users of (non-technical) services.
The \ac{NPS} results are in shown in~\autoref{fig:pred:E6nps}.
C1 and C3 (mainly passives) as well as C5, C6, and C8 (mainly detractors) achieve each a seemingly similar distribution while C4 stays between both groups.
This indicates that the \ac{NPS} is negatively affected if more \ac{LP} usage episodes are present.
Notably, also a saturation is indicated as C5 and C6 are seemingly similar.
However, the results contain outliers and the overall correlation coefficient of the \ac{NPS} with the final multi-episodic judgment is only~0.5.
\takeaway{Multi-episodic judgments alone do not suffice to predict service recommendations captured by the NPS. We therefore assume that the NPS is affected by additional factors.}

This highlights the need for future work to create holistic models capturing the overall service experience.

\section{Quality Prediction}

We now evaluate different approaches to predict the multi-episodic judgments based upon the episodic judgments.
Guse~et~al.~\cite{guse_modelling_2014, guse_multiepisodic_2017} proposed to predict a multi-episodic judgment by computing the \emph{weighted average} of prior episodic judgments.
The influence of each usage episode $e_i$ on the overall multi-episodic quality $m_n$ is expressed by its weight $a_i$.
This was found sufficient considering the amount of data and noise.
Using a weighted average also allows to account for a recency effect.
We evaluate the \emph{a)}~\ac{WF}\footnote{Selecting $\mathit{w := n}$, this model type becomes the average over all prior episodic judgments as proposed by \cite{moller_single-call_2011}.} and \emph{b)}~\ac{LW}.
Both functions are parametrized by the window parameter~$\mathit{w}$.
For \ac{WF}, it is limited to $\mathit{w}~\in~\mathbb{N}$ and $0~<~\mathit{w}~\leq~\mathit{n}$.
\begin{equation}\label{eq:weight:window}
 WF: a_i= \left\{
 \begin{array}{ll}
   1,& \text{if } i - n + w > 0 \\
   0,& \text{otherwise}
 \end{array}
 \right.
\end{equation}
\begin{equation}\label{eq:weight:linear}  %
 LW: a_i= \left\{
 \begin{array}{ll}
 	i - n + w,& \text{if } i - n + 2*w > 0 \\
   0,& \text{otherwise}
 \end{array}
 \right.
\end{equation}

We evaluate the prediction accuracy by computing the \ac{RMSD} of the episodic \ac{MOS} (\ie{}, input) and multi-episodic \ac{MOS} (\ie{}, output).

\begin{figure}[b]
	\centering
\begin{knitrout}
\definecolor{shadecolor}{rgb}{0.969, 0.969, 0.969}\color{fgcolor}
\includegraphics[width=\maxwidth]{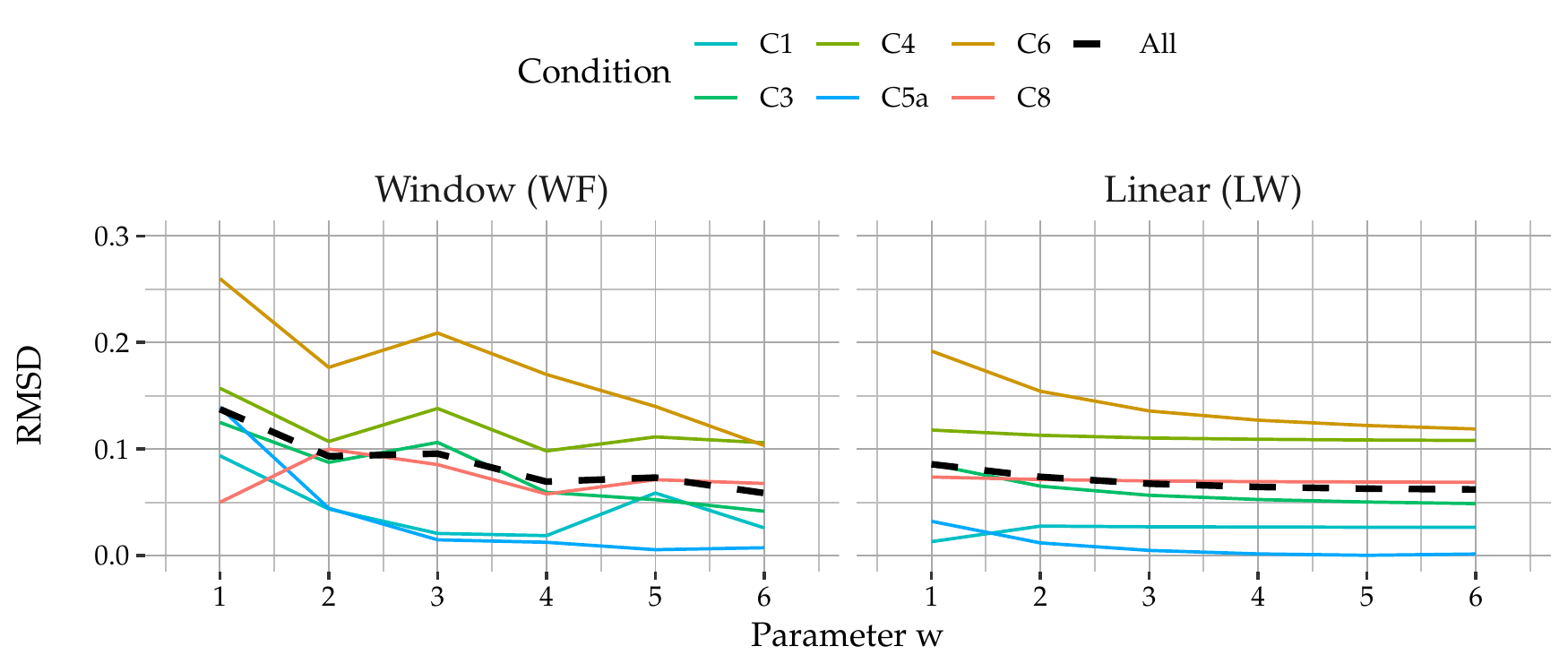} 

\end{knitrout}
    \caption{Multi-episodic prediction accuracy for C0.}
	\label{fig:pred:E6pred6}
\end{figure}

\afblock{Prediction C0 (HP only).}
The prediction accuracy improves for an increasing $\mathit{w}$~(\cf{}, \autoref{fig:pred:E6pred6}).
This is more prevalent for \ac{WF} than for \ac{LW}.
WF achieves its minimal \ac{RMSD} with $\mathit{w}=6$ (\ie, all prior episodes).
\ac{WF} provides only a marginal decrease for $\mathit{w}~\geq~3$.
\takeaway{The weighted average achieves a reasonable prediction accuracy while \ac{LW} provides a slightly better, robust performance.}
This is in line with \cite{guse_multiepisodic_2017}.

\begin{figure}
	\centering
\begin{knitrout}
\definecolor{shadecolor}{rgb}{0.969, 0.969, 0.969}\color{fgcolor}
\includegraphics[width=\maxwidth]{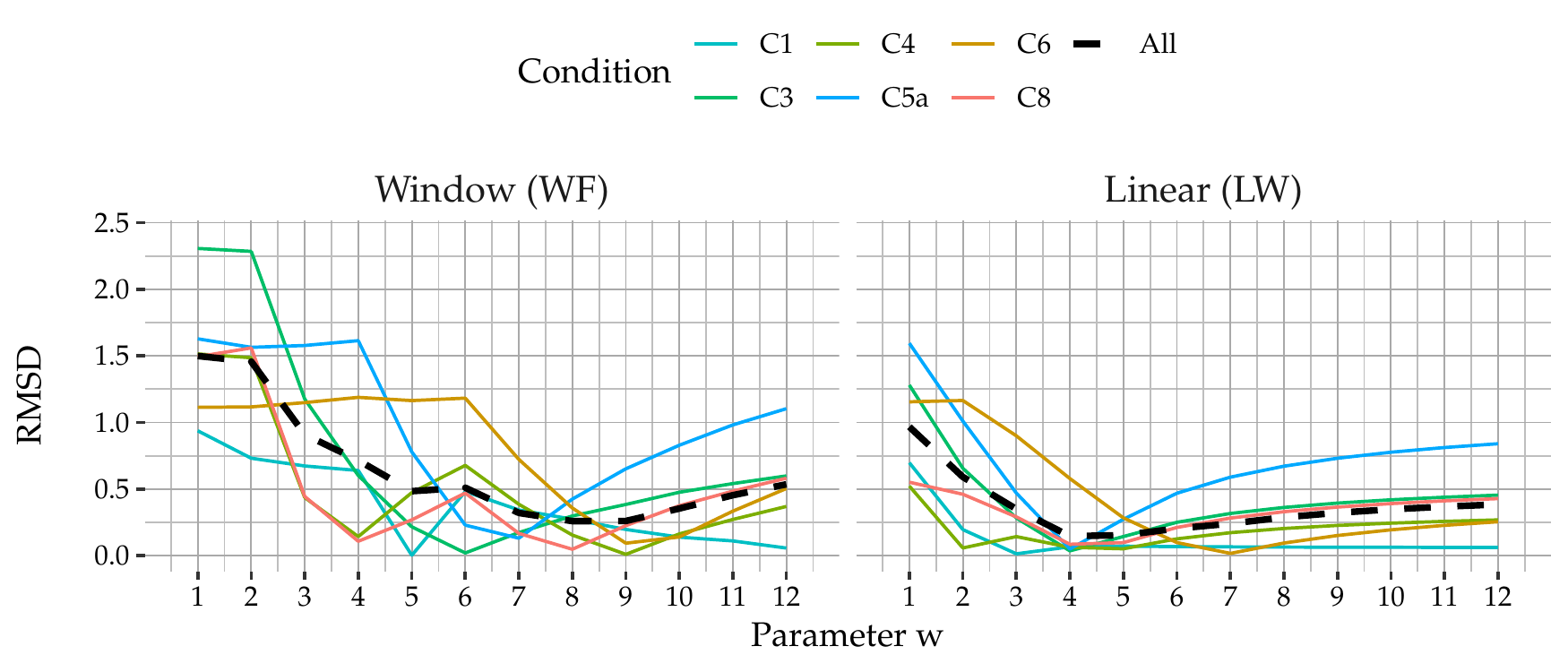} 

\end{knitrout}
	\caption{Multi-episodic prediction accuracy for all conditions (except C0).}
	\label{fig:pred:E6pred12}
\end{figure}

\afblock{Prediction C1-C8.}
With regard to the prediction of the multi\-/episodic judgment of the 6th~day, both weight functions perform differently~(\cf{}, \autoref{fig:pred:E6pred12}).
While LF reaches a minimal \ac{RMSD} at $\mathit{w}=4$ (\ie{},~0.15), WF not until $\mathit{w}=8$ (\ie{},~
0.26).
\takeaway{LF is preferable to WF, as a higher prediction accuracy is achieved. Also, LF requires a smaller $\mathit{w}$ while providing a higher robustness for choosing $\mathit{w}$.}

\begin{figure}[b]
	\centering
\begin{knitrout}
\definecolor{shadecolor}{rgb}{0.969, 0.969, 0.969}\color{fgcolor}
\includegraphics[width=\maxwidth]{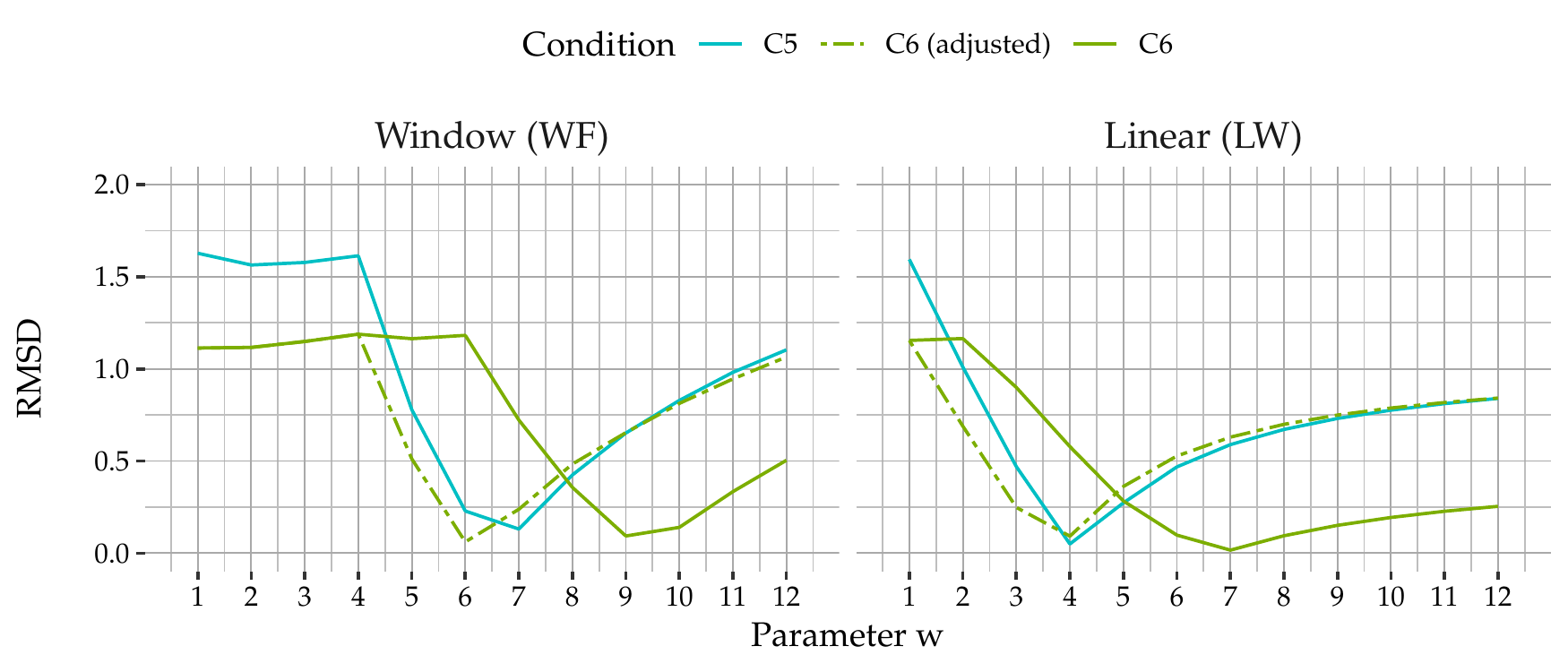} 

\end{knitrout}
	\caption{Multi-episodic prediction accuracy for saturation.}
	\label{fig:pred:SAT:E6}
\end{figure}

\afblock{Accounting for Saturation.}
In line with Guse~et.~al~\cite{guse_multiepisodic_2017}, a saturation could be observed although not taken into account for prediction.
For C6, the multi-episodic judgment remained at the same level as C5 although an additional usage episode was presented in \ac{LP}.
Also it remained approx. \unit[1]{pt} above the episodic judgment of \ac{LP}.
In fact, C5 and C6 only differ in the performance level of the 4th~usage episode.
As both were not judged differently, this suggests that this difference did not affect the formation process of the multi\-/episodic judgment.
In case of C6, we prepose to adjust the episodic judgments of the \ac{LP} usage episodes of the 4th~day by the average of the \ac{HP} usage episodes.
Then the window function can be applied without further modification.
For C6, this modification shifts the minimal \ac{RMSD} from $\mathit{w}=9$~to~ $\mathit{w}=6$ for WF and for LW from $\mathit{w}=7$ to $\mathit{w}=4$ (see \autoref{fig:pred:SAT:E6}).
Also, the prediction performance of C6~(adjusted) and C5 resemble each other closely.
\takeaway{This prefiltering approach for three consecutive \ac{LP} usage episodes allows to account for the saturation as it increases prediction performance.}

\section{Conclusion}
We presented the results of a multi-episodic experiment of an \ac{AoD} service.
The primary goal was to extend the work on multi-episodic perceived quality in one session \cite{guse_multiepisodic_2017} to several days.
For this reason, the same hypotheses and also a very similar experimental design was applied.
We made three observations.
First, increasing the number of \ac{LP} usage episodes decreases the directly following multi-episodic judgment (\hyponumber).
Here, the multi-episodic judgment reaches saturation showing that prior \ac{HP} usage episodes are still accounted for.
Second, we could not find a significant impact of a recency effect (\hypoposition) that was observed in prior work~\cite{guse_multiepisodic_2017} on single sessions.
Third, consecutive vs. non-consecutive presentation of \ac{LP} usage episode did not seem to affect multi-episodic judgments (\hypoconsecutive).
This is, interestingly, in line with multi-episodic experiments in one session \cite{guse_multiepisodic_2017} indicating that the time between usage episodes has a limited impact on the multi-episodic formation process.

\afblock{Future Work.}
Although our results are very promising, the formation process of multi-episodic perceived quality is still far from  well understood.
So far experiments forced participants to use service(s) in a certain manner by defining when and how to interact with them.
Thus, intentionally preventing variation in usage behavior and interaction.
However, in a normal setting a user has a motivation to interact with a service, sometimes a task, as well as a desired outcome and probably it's importance.
Therefore, (temporary) failures of services might prevent a user from fulfilling his/her task and therefore the multi-episodic formation process.
Also, the attribution of reduced service performance or failures might affect multi-episodic judgments (\eg{}, being in a remote area with limited mobile coverage).
Moreover, it is still open how different modalities (\eg{}, audio, video), service types (\eg{}, web~browsing), and  service bundles are actually judged and how the formation process is determined.

\bibliographystyle{IEEEtran}
\bibliography{Bibliography}

\begin{thebibliography}{10}
\providecommand{\url}[1]{#1}
\csname url@samestyle\endcsname
\providecommand{\newblock}{\relax}
\providecommand{\bibinfo}[2]{#2}
\providecommand{\BIBentrySTDinterwordspacing}{\spaceskip=0pt\relax}
\providecommand{\BIBentryALTinterwordstretchfactor}{4}
\providecommand{\BIBentryALTinterwordspacing}{\spaceskip=\fontdimen2\font plus
\BIBentryALTinterwordstretchfactor\fontdimen3\font minus
  \fontdimen4\font\relax}
\providecommand{\BIBforeignlanguage}[2]{{%
\expandafter\ifx\csname l@#1\endcsname\relax
\typeout{** WARNING: IEEEtran.bst: No hyphenation pattern has been}%
\typeout{** loaded for the language `#1'. Using the pattern for}%
\typeout{** the default language instead.}%
\else
\language=\csname l@#1\endcsname
\fi
#2}}
\providecommand{\BIBdecl}{\relax}
\BIBdecl

\bibitem{weiss_modeling_2009}
B.~Weiss, S.~Möller, A.~Raake, J.~Berger, and R.~Ullmann, ``Modeling call
  quality for time-varying transmission characteristics using simulated
  conversational structures,'' vol.~95, no.~6, pp. 1140--1151, 2009.

\bibitem{kahnemann_pain_1993}
D.~Kahneman, B.~L. Fredrickson, C.~A. Schreiber, and D.~A. Redelmeier, ``When
  more pain is preferred to less: Adding a better end,'' \emph{Psychological
  Science}, vol.~4, no.~6, pp. 401--405, Nov. 1993.

\bibitem{guse_multiepisodic_2017}
D.~Guse, B.~Weiss, F.~Haase, A.~Wunderlich, and S.~M\"{o}ller, ``Multi-episodic
  perceived quality for one~session of consecutive usage~episodes with a speech
  telephony service,'' \emph{Quality and User Experience}, vol.~2, no.~1, Aug.
  2017.

\bibitem{guse_modelling_2014}
D.~Guse, B.~Weiss, and S.~Möller, ``Modelling multi-episodic quality
  perception for different telecommunication services: first insights,'' in
  \emph{IEEE QoMEX}, 2014.

\bibitem{guse_duration_2016}
D.~Guse, A.~Wunderlich, B.~Weiss, and S.~Möller, ``Duration {{Neglect}} in
  {{Multi}}-episodic {{Perceived Quality}},'' in \emph{IEEE QoMEX}, 2016.

\bibitem{reichheld_one_2003}
F.~F. Reichheld, ``The {{One Number You Need}} to {{Grow}},'' \emph{Harvard
  Business Review}, Dec. 2003.

\bibitem{npscritism}
T.~Keiningham, B.~Cooil, T.~Andreassen, and L.~Aksoy, ``A longitudinal
  examination of net promoter and firm revenue growth,'' \emph{J Market},
  vol.~71, 2007.

\bibitem{moller_single-call_2011}
S.~Möller, C.~Bang, T.~Tamme, M.~Vaalgamaa, and B.~Weiss, ``From single-call
  to multi-call quality: a study on long-term quality integration in
  audio-visual speech communication,'' in \emph{INTERSPEECH}, 2011.

\bibitem{itu-t_recommendation_p.805_subjective_2007}
{ITU-T Recommendation P.805}, \emph{Subjective {{Evaluation}} of
  {{Conversational Quality}} (04/2007)}.

\bibitem{guse_macro-temporal_2013}
D.~Guse and S.~Möller, ``Macro-temporal {{Development}} of {{QoE}}: {{Impact}}
  of {{Varying Performance}} on {{QoE}} over {{Multiple Interactions}},'' in
  \emph{Proceedings of {{AIA-DAGA Conference}} on {{Acoustics}}}, 2013.

\bibitem{guse_thefragebogen_2019}
D.~Guse, H.~R. Orefice, G.~Reimers, and O.~Hohlfeld, ``{TheFragebogen}: A web
  browser-based questionnaire framework for scientific research,'' in
  \emph{IEEE QoMEX}, 2019.

\bibitem{itu-t_recommendation_p.851_subjective_2003}
{ITU-T Recommendation P.851}, \emph{Subjective quality evaluation of telephone
  services based on spoken dialogue systems (11/2003)}.

\bibitem{hossfeld2014qoe}
T.~Ho{\ss}feld, C.~Keimel, M.~Hirth, B.~Gardlo, J.~Habigt, K.~Diepold, and
  P.~Tran-Gia, ``{Best Practices for QoE Crowdtesting: QoE Assessment With
  Crowdsourcing},'' \emph{IEEE Transactions on Multimedia}, vol.~16, no.~2, pp.
  541--558, Feb 2014.

\end{thebibliography}

\end{document}